\documentclass[aps,prb,twocolumn,showpacs,superscriptaddress]{revtex4}
\usepackage{graphicx,color}
\usepackage{amsmath}
\usepackage{amssymb}
\usepackage{bm}

\begin{document}
\title{Interference of Cooper Pairs Emitted from Independent Superconductors}
\author{Mauro Iazzi}
\affiliation{International School for Advanced Studies (SISSA), via Beirut 2-4, I-34014 Trieste, Italy}
\author{Kazuya Yuasa}
\affiliation{Waseda Institute for Advanced Study, Waseda University, Tokyo 169-8050, Japan}

\date[]{April 9, 2010}

\begin{abstract}
We discuss the interference in the two-particle distribution of the electrons emitted from two independent superconductors.
It is clarified that, while the interference appearing in the antibunching correlation is due to the Hanbury Brown and Twiss effect, that in the positive correlation due to superconductivity is intrinsically different and is nothing but the first-order interference of Cooper pairs emitted from different sources.
This is the equivalent of the interference of two independent Bose-Einstein condensates.
\end{abstract}
\maketitle

A spectacular phenomenon in the physics of cold atomic gases is observed when two independently prepared Bose-Einstein condensates (BECs) are released and overlap, yielding interference fringes.\cite{ref:InterferenceBEC} 
Although particles emitted from independent sources do not normally exhibit Young-type (first-order) interference,\cite{ref:Loudon} two independent BECs do interfere.
The simplest explanation of this phenomenon is spontaneous symmetry breaking:\cite{ref:BECPitaevskiiStringari} 
the phases of the two gases are individually fixed upon condensation and their difference (relative phase) becomes well defined, which enables BECs to interfere like lasers.
Such a coherence is one of the distinguished features of condensed states.

The phenomenon of superconductivity is understandable as a kind of Bose-Einstein condensation.
Although electrons are fermions and a single electron can only occupy a single state, once \textit{Cooper pairs} are formed, they can condense like bosons as the system is cooled below a critical temperature, entering the superconducting phase.\cite{ref:Tinkham}
In the standard description of superconducting (BCS) states, Cooper pairs are condensed and the global $U(1)$ symmetry is spontaneously broken.\cite{ref:Tinkham}
A natural question now arises:
\textit{do Cooper pairs emitted from two independent superconductors interfere like BECs?}
\textit{Which kind of setup is required to reveal this interference?}
This is the main issue we address in this paper.
Interestingly, we shall see that \textit{first-order} interference of Cooper pairs is disclosed by \textit{two} detectors.

\begin{figure}[b]
\includegraphics[width=0.35\textwidth]{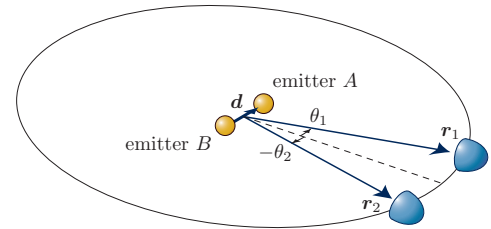}
\caption{(Color online) Electrons are emitted from two independent superconductors $A$ and $B$, with the phases of the superconducting states $\phi_A$ and $\phi_B$, respectively, and we count coincidences at $\bm{r}_1$ and $\bm{r}_2$ ($r_1=r_2=r$).
The emitting regions, both of size $w$, are separated by $\bm{d}$.
The directions to the detectors 1 and 2 are specified by the angles $\theta_1$ and $\theta_2$ measured from the dashed line perpendicular to $\bm{d}$.}
\label{fig:Setup}
\end{figure}

We consider field emission of electrons \cite{ref:OshimaNb} from two independent superconductors into vacuum (Fig.\ \ref{fig:Setup}).
Detectors do not detect Cooper pairs as a whole, but rather single electrons.
The electrons themselves are not condensed in the sources, so that there is no definite phase relationship between the electron wave functions originating from the two different sources, and no first-order interference like in a Young-type double-slit experiment.
Interference fringes are however found in the two-particle distribution (second-order interference \cite{ref:Loudon}).
As we shall see, two kinds of interference are present:  
one is due to the quantum exchange statistics, known as (the fermionic version of) the Hanbury Brown and Twiss (HBT) effect,\cite{ref:Loudon,ref:AntibunchingExps,ref:Antibunching-Kiesel,ref:SamuelssonNeder} while the other, intrinsically different, is clarified to be the \textit{first-order} interference of Cooper pairs emitted from different sources: the latter is the equivalent of the interference of two independent BECs.

We describe the emission of electrons from two superconductors $A$ and $B$ into vacuum \cite{ref:Samuelsson-SuperBell-cmt} by the Hamiltonian $H=H_A+H_B+H_V+V_A+V_B$,\cite{ref:TunnelingHamiltonian-SuperFE,ref:cooperbunch} where
\begin{align}
&
H_\ell=\sum_{\sigma=\uparrow,\downarrow}
\int d^3\bm{k}\,\omega_k\alpha_{\bm{k}\sigma}^{(\ell)\dag}\alpha_{\bm{k}\sigma}^{(\ell)},
\quad
\omega_k=\sqrt{\varepsilon_k^2+|\Delta|^2},
\label{eqn:HS}
\displaybreak[0]
\\
&
H_V=\sum_{\sigma=\uparrow,\downarrow}
\int d^3\bm{p}\,\varepsilon_pc_{\bm{p}\sigma}^\dag c_{\bm{p}\sigma},\quad\ \ \,%
\varepsilon_p=\frac{p^2}{2m}-\mu,
\displaybreak[0]\\
&
V_\ell= \sum_{\sigma=\uparrow,\downarrow}
\int d^3\bm{p}\,d^3\bm{k} 
\,T_{\bm{p}\bm{k}}^{(\ell)}
c_{\bm{p}\sigma}^\dag 
a_{\bm{k}\sigma}^{(\ell)}
+\text{H.c.}\ \ (\ell=A,B).
\label{eqn:V}
\end{align}
$H_{A/B}$ and $H_V$ are for the superconductors $A$/$B$ and the electrons in vacuum, respectively, while $V_{A/B}$ describe the emission from $A$/$B$ into vacuum. 
Coulomb repulsion among the emitted electrons becomes relevant only at much larger current densities than in typical experiments and can be neglected.\cite{ref:OshimaNb,ref:cooperbunch}
The operators $\alpha_{\bm{k}\sigma}^{(\ell)}$ (quasiparticle) and $a_{\bm{k}\sigma}^{(\ell)}$ (electron) are related by \cite{ref:Tinkham}
\begin{equation}
\begin{pmatrix}
\displaystyle
a_{\bm{k}\uparrow}^{(\ell)}\\
\displaystyle
a_{-\bm{k}\downarrow}^{(\ell)\dag}
\end{pmatrix}
=
\begin{pmatrix}
\smallskip
u_k&-v_ke^{i\phi_\ell}\\
v_ke^{-i\phi_\ell}&u_k
\end{pmatrix}
\begin{pmatrix}
\displaystyle
\alpha_{\bm{k}\uparrow}^{(\ell)}\\
\displaystyle
\alpha_{-\bm{k}\downarrow}^{(\ell)\dag}
\end{pmatrix}
\quad
(\ell=A,B),
\label{eqn:BogoliubovTr}
\end{equation}
with $u_k=\sqrt{(1+\varepsilon_k/\omega_k)/2}$ and $v_k=\sqrt{(1-\varepsilon_k/\omega_k)/2}$.
We set $\hbar=1$ and assume that the Fermi levels of the superconductors are the same, $\mu_A=\mu_B=\mu$, and the gap parameters $\Delta_{A/B}=|\Delta|e^{i\phi_{A/B}}$ are of the same magnitude but with different phases.
Note that the $U(1)$ symmetry of the system is broken on the basis of the standard BCS theory and the superconductors are endowed with constant phases $\phi_{A/B}$.\cite{ref:Tinkham}
The transmission matrices are
\begin{equation}
T_{\bm{p}\bm{k}}^{(A)}
= e^{-i(\bm{p}-\bm{k})\cdot\bm{d}/2}T_{\bm{p}\bm{k}},
\quad
T_{\bm{p}\bm{k}}^{(B)}
= e^{i(\bm{p}-\bm{k})\cdot\bm{d}/2}T_{\bm{p}\bm{k}},
\label{eqn:DisplacedT}
\end{equation}
where $T_{\bm{p}\bm{k}}
=\lambda h(\bm{p})\tilde{g}(\bm{p}-\bm{k})
$ with
$\tilde{g}(\bm{k})
=
e^{-k^2w^2/2}$,
$h(\bm{p})=e^{\varepsilon_p/2E_C}$: electrons are emitted from regions of radius $w$ with tunneling probability $\propto |h(\bm{p})|^2$, and 
the momentum transfer is ruled by $\tilde{g}(\bm{p}-\bm{k})$,\cite{ref:cooperbunch,ref:LateralEffects} with emitting regions displaced in space by $\pm\bm{d}/2$.
$E_C$ characterizes the low-energy cutoff of the tunneling probability and $\lambda$ the strength of the emission.

At time $t=0$, the emitters are in the BCS states with independently fixed phases $\phi_{A/B}$, while there is no electron outside.
Electrons then start to be emitted according to the Hamiltonian (\ref{eqn:HS})--(\ref{eqn:V}), and the emitted beam approaches a non-equilibrium steady state (NESS) \cite{ref:NESS_abbr2} at $t\to\infty$, in which we compute the normalized two-particle distribution at a given instant,\cite{ref:Loudon,ref:cooperbunch,ref:LateralEffects}
\begin{align}
Q(\bm{r}_1,\bm{r}_2) 
&= \frac{
\sum_{\sigma_1,\sigma_2}
\langle\psi_{\sigma_1}^\dag(\bm{r}_1)\psi_{\sigma_2}^\dag(\bm{r}_2)\psi_{\sigma_2}(\bm{r}_2)\psi_{\sigma_1}(\bm{r}_1)\rangle
}{
\sum_{\sigma_1,\sigma_2}
\langle\psi_{\sigma_1}^\dag(\bm{r}_1)\psi_{\sigma_1}(\bm{r}_1)\rangle
\langle\psi_{\sigma_2}^\dag(\bm{r}_2)\psi_{\sigma_2}(\bm{r}_2)\rangle
}
\nonumber
\displaybreak[0]
\\
&
=
1-\frac{
 |\gamma(\bm{r}_1, \bm{r}_2)|^2
-|\chi(\bm{r}_1,\bm{r}_2)|^2
}{2\gamma(\bm{r}_1,\bm{r}_1)\gamma(\bm{r}_2,\bm{r}_2)},
\label{eqn:Q}
\end{align}
where $\psi_\sigma(\bm{r})=\int d^3\bm{p}\,c_{\bm{p}\sigma}e^{i\bm{p}\cdot\bm{r}}/\sqrt{(2\pi)^3}$ is the field operator of the electrons in vacuum, and
\begin{align}
\gamma(\bm{r}_1,\bm{r}_2) 
&=\langle\psi_\uparrow^\dag(\bm{r}_1)\psi_\uparrow(\bm{r}_2)\rangle 
=\langle\psi_\downarrow^\dag(\bm{r}_1)\psi_\downarrow(\bm{r}_2)\rangle ,
\displaybreak[0]\\
\chi(\bm{r}_1,\bm{r}_2) 
&=\langle\psi_\uparrow(\bm{r}_1)\psi_\downarrow(\bm{r}_2)\rangle 
=-\langle\psi_\downarrow(\bm{r}_1)\psi_\uparrow(\bm{r}_2)\rangle.
\end{align}
Up to second order in $V_\ell$, both $\gamma$ and $\chi$ are given by the sum of two contributions, from emitters $A$ and $B$. 
The two contributions differ only for the phases of $T_{\bm{p}\bm{k}}^{(A/B)}$, due to the spatial displacement of the emitters, and for the phases of the superconductors $\phi_{A/B}$ in Bogoliubov's transformation (\ref{eqn:BogoliubovTr}). 
$\gamma$ and $\chi$ in the NESS ($t\to\infty$), at far places ($r\to\infty$), for $d/r,\,E_C/\mu,\,k_Fw^2/r,\,k_Fd^2/r,\,e^{-2k_F^2w^2}\ll1$ (with the Fermi momentum $k_F=2\pi/\lambda_F=\sqrt{2m\mu}$), are estimated to be \cite{ref:cooperbunch}
\begin{align}
\gamma(\bm{r}_1,\bm{r}_2) 
&\simeq2\cos[k_F(\hat{\bm{r}}_1-\hat{\bm{r}}_2)\cdot\bm{d}/2]
\gamma_0(\bm{r}_1,\bm{r}_2),
\label{eqn:GammaAna}
\displaybreak[0]\\
\chi(\bm{r}_1,\bm{r}_2) 
&\simeq2e^{i\bar{\phi}}
\cos[k_F(\hat{\bm{r}}_1+\hat{\bm{r}}_2)\cdot\bm{d}/2-\varDelta\phi/2]
\nonumber\\
&\qquad\qquad\qquad\qquad\qquad
{}\times
\chi_0(\bm{r}_1,\bm{r}_2,\bm{d}),
\label{eqn:ChiAna}
\end{align}
where
\begin{equation}
\gamma_0(\bm{r}_1,\bm{r}_2)
=Ae^{-4k_F^2w^2\sin^2[(\theta_1-\theta_2)/4]}
|\Delta|K_1(|\Delta|/E_C),
\end{equation}
\begin{widetext}
\begin{align}
&\chi_0(\bm{r}_1,\bm{r}_2,\bm{d})=\frac{\pi}{4i}A|\Delta|e^{2ik_Fr}
\nonumber\\
&\qquad\qquad\qquad\qquad
\times
\biggl[
e^{-4k_F^2w^2\sin^2[(\pi-|\theta_1-\theta_2|)/4]}
e^{-[(\hat{\bm{r}}_1-\hat{\bm{r}}_2)\cdot\bm{d}]^2/16\pi^2\xi^2
-ir/2\pi^2k_F\xi^2
}
H_0^{(2)}\!\left(
\tfrac{iw^2}{\pi^2\xi^2}
+\tfrac{i[(\hat{\bm{r}}_1-\hat{\bm{r}}_2)\cdot\bm{d}]^2}{16\pi^2\xi^2}
-\tfrac{r}{2\pi^2k_F\xi^2}
\right)
\nonumber
\\
&\qquad\qquad\qquad\qquad\qquad\qquad\qquad\qquad
{}-\frac{
4/\pi
}{\sqrt{ ir/k_Fw^2}}
e^{-k_F^2w^2\sin^2[(\pi-\theta_1+\theta_2)/2]}
e^{ik_F[(\hat{\bm{r}}_1-\hat{\bm{r}}_2)\cdot\bm{d}]^2/16r}
\biggr] \qquad (|\theta_1-\theta_2|\le\pi),
\label{eqn:ChiD}
\end{align}
\end{widetext}
$K_1(z)$ and $H_0^{(2)}(z)$ are Bessel and Hankel functions, respectively, and $A=2\pi^2mk_F\lambda^2/w^2r^2$. 
Pippard's length $\xi=k_F/\pi m|\Delta|$ characterizes the extension of a Cooper pair in the superconductors,\cite{ref:Tinkham}
$
\varDelta\phi=\phi_A-\phi_B
$, and $
\bar\phi=(\phi_A+\phi_B)/2
$.

\begin{figure}[t]
\includegraphics[width=0.48\textwidth]{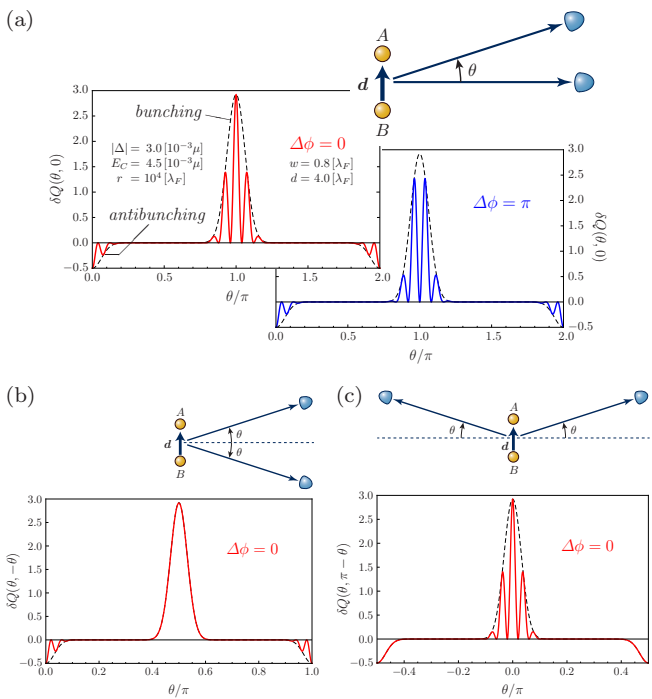}
\caption{(Color online) Correlation function $\delta Q(\theta_1,\theta_2)=Q-1$ for different placements of the detectors. The interference pattern in the bunching correlation (indicated by ``\textit{bunching}'') is affected by the relative phase $\varDelta\phi$, while that in the antibunching (indicated by ``\textit{antibunching}'') is not.}
\label{fig:Configs}
\end{figure}
As discussed in Ref.\ \onlinecite{ref:cooperbunch} for the single-emitter case, $\gamma$ describes the one-particle state of the emitted electrons, while $\chi$ represents the pair correlation. 
$\gamma$ exhibits correlations for $\theta_1-\theta_2\sim0$, while $\chi$ for $\theta_1-\theta_2\sim\pi$. 
As is clear from (\ref{eqn:Q}), $\gamma$ reduces the coincidence counts for electrons detected in collinear directions (antibunching), while $\chi$ enhances it for electrons emitted in opposite directions (bunching).
Notice that $\chi$ is absent for normal emitters. 
It describes the emission of a Cooper pair (Andreev process), and the correlation at $\theta_1-\theta_2\sim\pi$ reflects the Cooper pairing of momenta $\bm{k}$ and $-\bm{k}$: the two electrons of a Cooper pair are emitted in opposite directions.
Since a Cooper pair is spin-singlet, its wave function is symmetric in space, yielding positive correlation.\cite{ref:BuettikerReview2000}

In the case of the double emitters, interference fringes appear in the antibunching envelope of $\gamma_0$ and in the bunching one of $\chi_0$: the $A$ and $B$ contributions interfere.
See (\ref{eqn:GammaAna})--(\ref{eqn:ChiAna}) and Fig.\ \ref{fig:Configs}.
The fringe spacing is given by
$\delta\theta_i=2\pi/k_Fd$ (for $\theta_i\sim0$, $i=1,2$), while the widths of the envelopes are $\varDelta\theta\sim1/k_Fw$. 
Notice that \textit{no interference} is observed in the one-particle distribution $\propto\gamma(\bm{r},\bm{r})
=2A|\Delta|K_1(|\Delta|/E_C)
$ (no first-order interference), since the two emitters are \textit{independent}.

Observe now that the interference fringes in the antibunching and bunching correlations behave in different ways, due to the different signs of $k_F(\hat{\bm{r}}_1\mp\hat{\bm{r}}_2)\cdot\bm{d}/2$ in (\ref{eqn:GammaAna})--(\ref{eqn:ChiAna}).
In fact, the bunching profile does not oscillate when the detectors are moved according to $\theta_1=-\theta_2$ [Fig.\ \ref{fig:Configs}(b)], while no fringes show up in the antibunching correlation for $\theta_1=\pi-\theta_2$ [Fig.\ \ref{fig:Configs}(c)].
There is another remarkable difference between the interferences in the bunching and antibunching correlations: the former interference is affected by the phases of the two superconductors, as can be clearly seen in Fig.\ \ref{fig:Configs}(a).

\begin{figure}[t]
\includegraphics[width=0.48\textwidth]{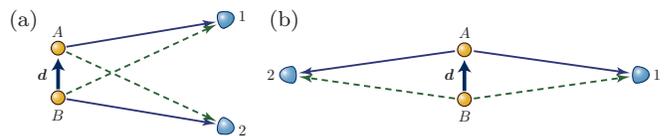}
\caption{(Color online) Interfering processes (solid lines vs dashed) giving rise to the fringes in the antibunching correlation (a) and to those in the bunching one (b).}
\label{fig:Paths}
\end{figure}

These differences suggest that the interference in the antibunching and that in the bunching have different physical origins.
First of all, the latter interference is absent for normal emitters, while the former is essentially irrelevant to superconductivity.
The former is due to the HBT effect and is a consequence of the superposition of the following two alternatives: 
in one case an electron goes from emitter $A$ to detector 1 and the other one from $B$ to 2, while in the other case they travel from $A$ to 2 and from $B$ to 1 [Fig.\ \ref{fig:Paths}(a)].
These two processes with the destinations interchanged are indistinguishable and are to be superposed according to the Fermi statistics, resulting in the interference in the antibunching correlation.
Actually, in the far-field regime, the two-particle wave functions for these two alternatives differ only for the phases, i.e., $e^{-ik_F\hat{\bm{r}}_1\cdot\bm{d}/2}e^{ik_F\hat{\bm{r}}_2\cdot\bm{d}/2}$ for the former case and $e^{-ik_F\hat{\bm{r}}_2\cdot\bm{d}/2}e^{ik_F\hat{\bm{r}}_1\cdot\bm{d}/2}$ for the latter,\cite{note:ShiftWave} whose superposition yields the interference term $\sim\cos[k_F(\hat{\bm{r}}_1-\hat{\bm{r}}_2)\cdot\bm{d}]$, which is the origin of the oscillation in $|\gamma(\bm{r}_1,\bm{r}_2)|^2$ [see (\ref{eqn:GammaAna})].
The interference fringes in the bunching correlation, on the other hand, is due to the superposition of different alternatives.
Look at the correlation function $\chi$ in (\ref{eqn:ChiAna}). 
It is the superposition of the following two contributions:
\begin{align}
\chi(\bm{r}_1,\bm{r}_2)
={}&e^{i\phi_A}
e^{-ik_F(\hat{\bm{r}}_1+\hat{\bm{r}}_2)\cdot\bm{d}/2}
\chi_0(\bm{r}_1,\bm{r}_2,\bm{d})
\nonumber\\
&{}
+e^{i\phi_B}
e^{ik_F(\hat{\bm{r}}_1+\hat{\bm{r}}_2)\cdot\bm{d}/2}
\chi_0(\bm{r}_1,\bm{r}_2,-\bm{d}),
\label{eqn:SuperpositionCooper}
\end{align}
i.e., both electrons originate from a Cooper pair in $A$ in the first term, while both from a Cooper pair in $B$ in the second [Fig.\ \ref{fig:Paths}(b)]. 
This is different from the superposition giving rise to the HBT effect [Fig.\ \ref{fig:Paths}(a)].
Instead, Eq.\ (\ref{eqn:SuperpositionCooper}) is the superposition of \textit{the wave functions of a single Cooper pair} originating from $A$ and $B$. 
Equation (\ref{eqn:SuperpositionCooper}) is \textit{the first-order interference of a Cooper pair from independent sources} (whose phases are well defined) and is \textit{the equivalent of the interference of BECs}. 
Notice that, while the HBT effect is insensitive to the phases of the two sources, the interference of Cooper pairs is sensitive to them.
Cooper pairs from different sources can exhibit interference, thanks to the fixed phases of the BCS states due to symmetry breaking, exactly like in the case of the interference of BECs.
Interestingly, in order to probe this \textit{first-order} interference, \textit{two} detectors are required.
Each Cooper pair is not detected as a whole by a single detector, but is captured by a pair of coincident detections in opposite directions.
The first-order interference of Cooper pairs is disclosed by the detections of single electrons by one detector under the condition that the other detector in the opposite direction clicks. 

The interference of Cooper pairs from independent sources bears the following interesting features.
Contrary to the case of a superconducting quantum interference device (SQUID), the phases of the two superconductors are not fixed by an external parameter, e.g., by a magnetic field, since a Josephson current, which would tend to align the phases, is absent. 
This assumption is reasonable, since the strong electric field drives the electrons away from the emitters and there is no current flowing between the two superconductors.
As a result, each time the system undergoes the superconducting phase transition, the relative phase $\varDelta\phi$ assumes a different value varying randomly from $0$ to $2\pi$.
Such a change would be observed by repeatedly breaking the superconducting state, e.g., by heating or passing a high current through one of the tips and waiting for a new realization of the superconducting state.\cite{ref:OshimaNb}
In an actual experiment, the spatial distribution of the emitted electrons may be distorted, since the extracting field is not likely spherically symmetric, or the surface of the tip may not be smooth. 
For these reasons, the contributions of the bunching and antibunching effects may overlap. 
In such a case, the effects due to Cooper pairs are detected by repeatedly breaking the superconducting state.
Any change in the correlation pattern after the recovery of the superconducting state is attributed to the Cooper-pair correlation.
Another interesting change in the interference pattern can also be observed.
As the emission goes on, new electrons are supplied to the superconductors.
They are not paired to the emitted ones and newly formed Cooper pairs may assume different phases.
The experiments of the type discussed here may detect such evolution of the phase of the superconductor.

We finally comment on the feasibility of the present scheme.
The Young-type (first-order) interference of electrons in vacuum has been observed with fairly good visibility in various field-emission experiments,\cite{ref:Tonomura-AJP1989,ref:SpaceCoherence} proving that decoherence during propagation in vacuum is negligible.
A coincidence experiment has recently been carried out with electrons field-emitted from a single nonsuperconducting tip, to test the antibunching (second-order interference) of electrons in vacuum.\cite{ref:Antibunching-Kiesel}
In addition, field emission from a superconductor has been reported in Ref.\ \onlinecite{ref:OshimaNb}.
All these experimental achievements support the feasibility of the present scheme.
Typical parameters in the ordinary field-emission experiments are $w\sim50\,\text{nm}$, $r\sim10\,\text{cm}$, $\lambda_F\sim0.3\,\text{nm}$, and $E_C\sim0.1\,\text{eV}$. For a gap parameter $|\Delta|\sim10^{-3}\,\text{eV}$, the height of the positive peak is estimated to be $\delta Q\sim10^{-4}$.
The angular width of the positive peak is $\varDelta\theta\sim10^{-3}$, while the angular fringe spacing is $\delta\theta\sim10^{-4}$ for $d\sim1\,\mu\text{m}$: magnification of the electron beams \cite{ref:Antibunching-Kiesel,ref:SpaceCoherence} by a factor $\sim10^2$ leads to the fringe spacing $\sim1\,\text{mm}$.
The decay of the positive correlation in the delay time $\tau$ is slow like $\sim|\Delta\tau|^{-1}$,\cite{ref:cooperbunch} and a long coherence time is expected, which is of order $\sim|\Delta|^{-1}\sim1\,\text{ps}$ for the above gap parameter and is a bit shorter than the response time $\sim10\,\text{ps}$ of the detector employed in the coincidence experiment in Ref.\ \onlinecite{ref:Antibunching-Kiesel}.
The detector resolution in field-emission measurements is steadily improving, both in space \cite{ref:SpaceCoherence} and time,\cite{ref:TimeCoherence} and various kinds of  strong superconductors have been discovered these years, which  would provide a better experimental configuration.
Note that the strength of the positive correlation is roughly proportional to $\delta Q\propto|\Delta|^2/E_C^2$.

In summary, we have discussed the interference in the two-particle distribution of the electrons field-emitted from two independent superconductors.
Interference fringes appear in the antibunching and bunching correlations.
It has been clarified that the latter interference is intrinsically different from the HBT effect and is nothing but the first-order interference of Cooper pairs from independent sources.
This is the equivalent of the interference of independent BECs.
This interference appears with definite fringe spacing but with different offsets for different realizations of the superconducting states.

The above arguments are based on the idea of the BCS states with fixed phases. 
It is worth recalling however that interference between BECs shows up even if the number of the atoms in each cloud is fixed and finite, i.e., in the absence of symmetry breaking.
The relative phase is established by the measurements of the positions of the atoms in the clouds and interference fringes are found at each snapshot.\cite{ref:BECInt2}
It would be interesting to explore such a possibility and explain the appearance of the phases of the superconductors without resorting to the concept of the spontaneous symmetry breaking.

\acknowledgments
We thank C. Oshima for discussions.
This work is supported  
by a Special Coordination Fund for Promoting Science and Technology and the Grant-in-Aid for Young Scientists (B) both from MEXT, Japan,
by the bilateral Italian-Japanese Projects of MUR, Italy, and by the Joint Italian-Japanese Laboratory of MAE, Italy.


\begin{thebibliography}{10}

\bibitem{ref:InterferenceBEC}
M.~R. Andrews, C.~G. Townsend, H.-J. Miesner, D.~S. Durfee, D.~M. Kurn, and W.
  Ketterle, Science \textbf{275},  637  (1997).

\bibitem{ref:Loudon}
R. Loudon, \textit{The Quantum Theory of Light}, 3rd ed. (Oxford University
  Press, Oxford, 2000).

\bibitem{ref:BECPitaevskiiStringari}
L. Pitaevskii and S. Stringari, \textit{Bose-Einstein Condensation} (Oxford
  University Press, Oxford, 2003).

\bibitem{ref:Tinkham}
M. Tinkham, \textit{Introduction to Superconductivity}, 2nd ed. (Dover
  Publications, New York, 1996).

\bibitem{ref:OshimaNb}
K. Nagaoka, T. Yamashita, S. Uchiyama, M. Yamada, H. Fujii, and C. Oshima,
  Nature (London) \textbf{396},  557  (1998).

\bibitem{ref:AntibunchingExps}
R.~C. Liu, B. Odom, Y. Yamamoto, and S. Tarucha, Nature (London) \textbf{391}, 263 (1998); 
M. Henny, S. Oberholzer, C. Strunk, T. Heinzel, K. Ensslin, M. Holland, and C.
  Sch\"onenberger, Science \textbf{284}, 296 (1999); 
W.~D. Oliver, J. Kim, R.~C. Liu, and Y. Yamamoto,
  \textit{ibid.} \textbf{284}, 299 (1999); 
M. Iannuzzi, A. Orecchini, F. Sacchetti, P. Facchi, and S. Pascazio, Phys. Rev.
  Lett. \textbf{96}, 080402 (2006); 
T. Rom, {Th. Best}, D. van Oosten, U. Schneider, S. F\"olling, B. Paredes, and
  I. Bloch, Nature (London)
  \textbf{444}, 733 (2006); 
T. Jeltes, J.~M. McNamara, W. Hogervorst, W. Vassen, V. Krachmalnicoff, M.
  Schellekens, A. Perrin, H. Chang, D. Boiron, A. Aspect, and C.~I. Westbrook, \textit{ibid.}
  \textbf{445}, 402 (2007).

\bibitem{ref:Antibunching-Kiesel}
H. Kiesel, A. Renz, and F. Hasselbach, Nature (London) \textbf{418},  392
  (2002).

\bibitem{ref:SamuelssonNeder}
P. Samuelsson, E.~V. Sukhorukov, and M. B\"uttiker, Phys. Rev. Lett.
  \textbf{92}, 026805 (2004);
I. Neder, N. Ofek, Y. Chung, M. Heiblum, D. Mahalu, and V. Umansky, Nature (London)
  \textbf{448}, 333 (2007).

\bibitem{ref:Samuelsson-SuperBell-cmt}
Electron emission from a single superconductor to normal leads via two contacts
  is considered to generate orbitally entangled electrons in P. Samuelsson,
  E.~V. Sukhorukov, and M. B\"uttiker, Phys. Rev. Lett. \textbf{91}, 157002
  (2003).

\bibitem{ref:TunnelingHamiltonian-SuperFE}
J. Bardeen, Phys. Rev. Lett. \textbf{6}, 57 (1961); \textbf{9}, 147 (1962); M.~H. Cohen, L.~M. Falicov, and J.~C. Phillips, \textit{ibid.} \textbf{8}, 316 (1962); J.~W. Gadzuk, Surf. Sci. \textbf{15},
  466 (1969).

\bibitem{ref:cooperbunch}
K. Yuasa, P. Facchi, R. Fazio, H. Nakazato, I. Ohba, S. Pascazio, and S.
  Tasaki, Phys. Rev. B \textbf{79}, 180503(R) (2009); K. Yuasa,
  \textit{ibid.} \textbf{80}, 104516 (2009).

\bibitem{ref:LateralEffects}
K. Yuasa, P. Facchi, H. Nakazato, I. Ohba, S. Pascazio, and S. Tasaki, Phys. Rev. A \textbf{77},  043623  (2008).

\bibitem{ref:NESS_abbr2}
W. Aschbacher, V. Jak\v{s}i\'c, Y. Pautrat, and C.-A. Pillet, in \textit{Open Quantum Systems III}, edited by
  S. Attal, A. Joye, and C.-A. Pillet (Springer, Berlin, 2006), pp.\ 1--66; S.
  Tasaki and J. Takahashi, Prog. Theor. Phys. Suppl. \textbf{165}, 57 (2006).

\bibitem{ref:BuettikerReview2000}
{Ya. M. Blanter} and M. B\"uttiker, Phys. Rep. \textbf{336},  1  (2000).

\bibitem{note:ShiftWave}
The electron going to detector $i\,(=1,2)$ propagates with momentum
  $k_F\hat{\bm{r}}_i$ in the far-field regime, and the wave function shifted by
  $\pm\bm{d}/2$ gains a phase $e^{\mp ik_F\hat{\bm{r}}\cdot\bm{d}/2}$,
  depending on its origin, emitter $A$ (upper sign) or $B$ (lower).

\bibitem{ref:Tonomura-AJP1989}
A. Tonomura, J. Endo, T. Matsuda, T. Kawasaki, and H. Ezawa, Am. J. Phys. \textbf{57},  117  (1989).

\bibitem{ref:SpaceCoherence}
B. Cho, T. Ichimura, R. Shimizu, and C. Oshima, Phys. Rev. Lett. \textbf{92}, 246103 (2004); T. Ishikawa, B. Cho, E. Rokuta, and C. Oshima, Appl. Phys. Express \textbf{1}, 077001 (2008).

\bibitem{ref:TimeCoherence}
B. Cho, T. Itagaki, and C. Oshima, Appl. Phys. Lett. \textbf{91},  051916
  (2007).

\bibitem{ref:BECInt2}
J. Javanainen and S.~M. Yoo, Phys. Rev. Lett. \textbf{76}, 161 (1996); 
J.~I. Cirac, C.~W. Gardiner, M. Naraschewski, and P. Zoller, Phys. Rev. A \textbf{54}, R3714 (1996); Y. Castin and
  J. Dalibard, \textit{ibid.} \textbf{55}, 4330 (1997); A. Polkovnikov, E.
  Altman, and E. Demler, Proc. Natl. Acad. Sci. USA \textbf{103}, 6125 (2006).

\end{thebibliography}
\end{document}